\documentclass{elsart}
\usepackage[english]{babel}
\usepackage{amssymb}
\usepackage{amsfonts}
\usepackage{graphicx}

\newcommand{\bra}[1]{\langle #1|}
\newcommand{\ket}[1]{|#1\rangle}

\begin{document}

\begin{frontmatter}

\title{On the existence of black hole evaporation yet again}

\author{V.A. Belinski}
\ead{belinski@icra.it}

\address{INFN and ICRA, Rome University ``La Sapienza", 00185 Rome, Italy, and IHES, F--91440 Bures-sur Yvette, France}

\begin{abstract}
A new argument is presented confirming the point of view that a Schwarzshild black hole formed during a collapse process does not radiate.
\end{abstract}


\end{frontmatter}

\section{Introduction}

Articles \cite{1}--\cite{7} showed the absence of a theoretical basis for the existence of the so-called Unruh effect and in this context we have been asked by many researchers whether or not these results have some consequences for the black hole evaporation effect \cite{8} as well. The question is quite pertinent because these effects are related and some authors have even treated them as essentially the same phenomenon. I hope to clarify this point here.         

In 1995 I had already stated the impossibility of any quantum evaporation process taking place in the Schwarzschild black hole spacetime \cite{9}. Now I can repeat the same assertion with even more confidence because the work in  \cite{1}--\cite{7} indeed allows one to derive additional support in favour of the groundlessness of the conventional picture.  These new arguments together with the fact that in the field of a black hole created by gravitational collapse, there does not exist any quasiclassical sub-barrier motion corresponding to the particle creation process, again confirm my claim.

Section 3 shows the absence of the aforementioned tunnel transition directly for the Schwarzschild case (unlike  in \cite{9} which used the analogue of a black hole in the dilaton gravity model). Section 4 shows that the second quantization procedure from which the conventional evaporation picture follows is inapplicable to the problem at hand, because in the framework of this procedure the initial state of the quantum field in principle cannot be chosen to be the Minkowski vacuum. In fact the initial state in the traditional approach should be interpreted as a condensate which consists of an infinite number of pairs of correlated particles falling from spatial infinity mainly along  the ``last ray" (the continuation of the horizon into the past). Namely these particles which already existed at the onset of collapse leak out continuously into the external space at late times which can create an illusion of radiation produced by the black hole. Such an exotic process is due to the divergence of the particle number and energy densities  on the  last ray. Due to this infinite reserve of the energy in the initial state\footnote{Clearly such singular classical matter  cannot be realized in Minkowski spacetime without destructive back-reaction to the geometry and so the physical sense of the conventional picture remains questionable even after an adequate interpretation of its quantization procedure.} the stationary energy flux can arise also at the last stages of collapse in spite of the infinite red shift. Thus there is no process of late time particle production and at the end of the collapse, one observes only those particles which have already been stored in the system from the beginning. The same is true with regard to the thermal spectrum of the emergent flux: this spectrum is a direct result of the statistical noise \cite{10} which is present in the system from the beginning due to the peculiarity of the initial conditions. There are no paradoxes with the evolution of the pure state into a mixed one.

It goes without saying that there is no reason to choose such an extraordinary initial state for the quantum field. Although a rigorous solution of the real problem with the standard Minkowski vacuum as the initial state is still not within our grasp, the qualitative character of such solution can be predicted easily. The Minkowski vacuum is a translationally invariant state, therefore any quantization scheme should not give a physical preference to the last ray with respect to other locations at the initial stages of the collapse. From this it follows that the condition of translational invariance of the vacuum requires that any quantization be unitarily equivalent to the one which uses the modes which are regular on the last ray.\footnote{Quantization of singular modes in this case needs special care. Namely an ill-treatment of singular modes in the traditional picture leads to the fact that the initial state cannot be chosen to be translationally invariant.} However, in the outer space such modes are going to die away exponentially as the Schwarzschild time tends to infinity. Hence it is impossible to have a stationary radiation flux in these modes. Naturally, during the collapse, as in any non-stationary process, some amount of particles will be created but during the late phases this process will start to damp out and in the asymptotic future will disappear completely. 

It is worth remarking that a black hole's  inability to radiate energy makes it a peculiar object with respect to thermodynamics. A body which can only absorb and not radiate cannot reach an equilibrium with the environment, so that in principle thermodynamics cannot be applied to black holes. In particular,  no physically sensible notions of temperature and entropy can be defined for them.  This is natural because black holes have no intrinsic degrees of freedom needed for the statistical approach to be applied.

\section{The model}

It is well known that all qualitative characteristics of the assumed evaporation effect can already be seen in two-dimensional spacetime. Moreover, without loss of generality one can model the collapsing body by an infinitely thin light spherical shell. Outside the shell one has the Schwarzschild metric:
\begin{eqnarray}
\label{intout}
-(ds^2)_{out}=-\left(1-\frac{r_g}{r}\right)c^2dt^2+\left(1-\frac{r_g}{r}\right)^{-1}dr^2
\end{eqnarray}
and inside just flat spacetime:
\begin{eqnarray}
\label{intin}
	-(ds^2)_{in}=-c^2d\tau^2+dr^2 \,,
\end{eqnarray}
with the same spatial coordinate $r$ which is continuous across the shell. Using the tortoise variable $r^*$:
\begin{eqnarray}
	r^*=r+r_g\ln\left(\frac{r}{r_g}-1\right) \,,
\end{eqnarray}
the equation of the motion of the shell can be written as
\begin{eqnarray}
\label{eqm}
	r^*=2r_g-ct \,.
\end{eqnarray}
Matching the metrics (\ref{intout}) and (\ref{intin}) at the shell we get the following correspondence between the times $t$ and $\tau$:
\begin{eqnarray}
\label{corr}
	ct=c\tau-r_g\ln\left(1-\frac{c\tau}{r_g}\right) \,.
\end{eqnarray}
The arbitrary integration constants in Eqs.~(\ref{eqm}) and (\ref{corr}) have been chosen to have special values only for convenience.

\begin{figure}[ht]
	\begin{center}
		\includegraphics[width=3in]{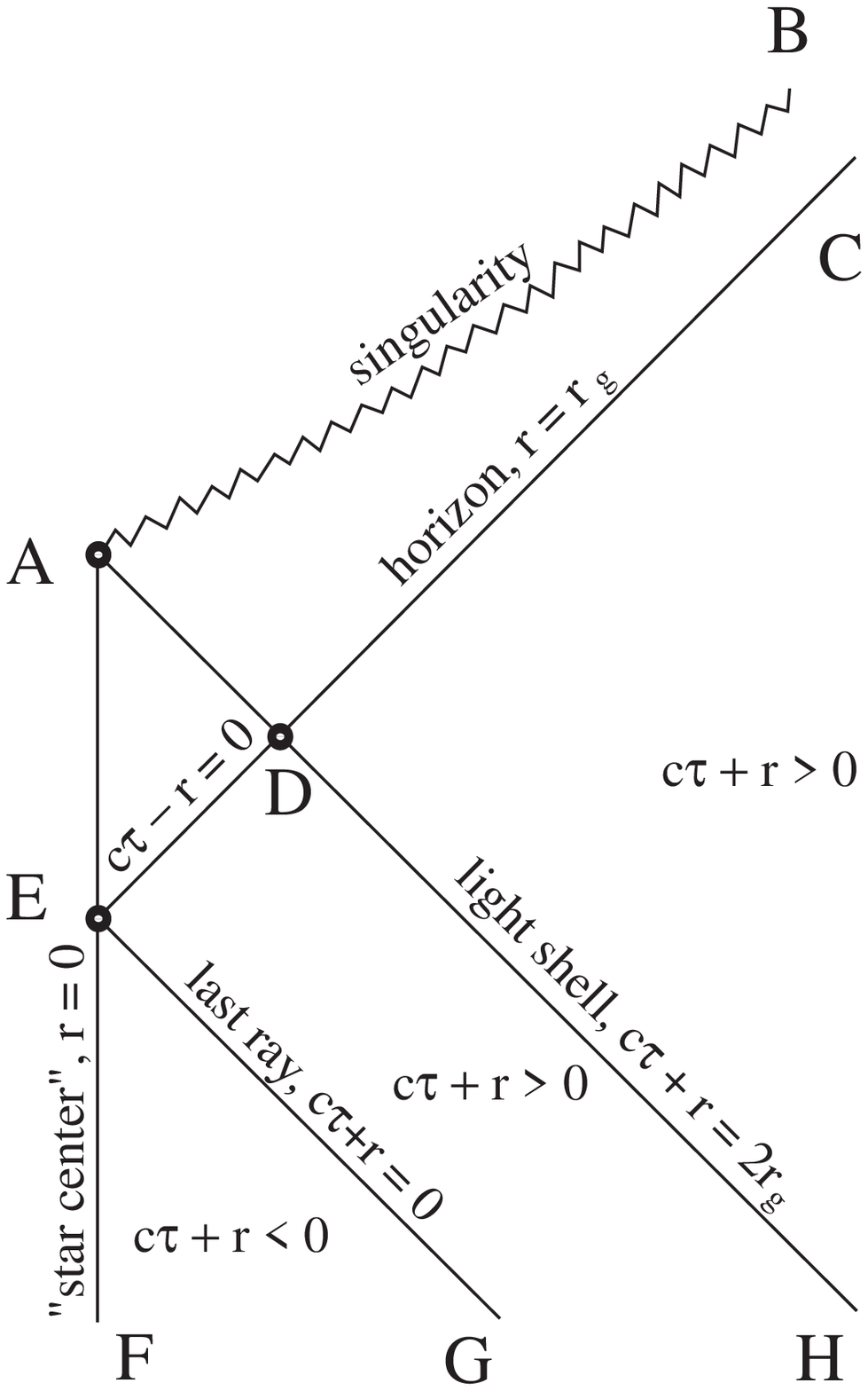}
	\end{center}
	\label{fig:1}
	\caption{}
\end{figure}

The collapse diagram is shown in Fig.~1. Everywhere outside the shell (namely in the region BADH of Fig.~1) we can use Kruskal coordinates $\eta$ and $\zeta$:
\begin{eqnarray}
\label{Krucoords}
	-ds^2=\frac{4r_g^3}{r}e^{-\frac{r}{r_g}}(-d\eta^2+d\zeta^2) \,, \quad \quad
	\eta^2-\zeta^2=\left(1-\frac{r}{r_g}\right)e^{\frac{r}{r_g}} \,,
\end{eqnarray}
which are related to the Schwarzschild coordinates in the following way:
\begin{eqnarray}
	r^*+ct=2r_g\ln(\zeta+\eta) \,, \quad\quad r^*-ct=2r_g\ln(\zeta-\eta) \,.
\end{eqnarray}
And finally,  in analyzing the second quantization procedure (section 4), we will restrict ourselves to a massless scalar field because the phenomenon under investigation is supported by extremely high frequencies and in this energy domain the influence of the mass can be neglected.

\section{Quasiclassical analysis}

In the traditional approach the matter of the collapsing body does not interact directly with the quantum field and in this sense gives no direct contribution to the evaporation effect. Furthermore the history of the collapse is of no importance. The radiation responsible for evaporation is assumed to be due to the particles created by the gravitational field of an already formed black hole at late times in the Schwarzschild future. A prevalent popular explanation is that one of the particles appears inside the horizon (in region ABCD, Fig.~1) and other outside of it (in region CDH). The first one is unable to move to the region outside the horizon and falls down into the singularity, while the second particle escapes to infinity to be detected there by the distant observer. It only makes sense to assume such an arrangement of the creation points  since it is well known that outside the horizon of a static black hole, no particle pairs can be created,\footnote{In this region no particle trajectory can cross the barrier between the seas of  positive and negative Schwarzschild energy states. The situation is qualitatively the same as for the free field in flat spacetime.} while the case where both particles are created inside horizon is of no interest because neither can escape to the external space. 

In fact it is easy to see that no pair can be created in the gravitational field of the black hole formed during the collapse independently of the mutual disposition of the assumed creation points. Analysis presented below shows that a particle in the external field described by the metric (\ref{Krucoords}) can perform the tunneling from some point ($\eta$,$\zeta$) only to the point ($-\eta$,$-\zeta$),  that is the creation points should be reflections of each other across the origin.  However, in the domain of interest, that is in the empty space outside the collapsing matter (region BADH, Fig.~1), no such pair of points exists.

Let us first see how the quasiclassical particle creation process can be described in a non-gravitational setting. When a particle moves in some external field in the flat spacetime of the metric $-ds^2=-c^2dt^2+dx^2$, we always have the following integral of the motion:
\begin{eqnarray}
	\left(\frac{cdt}{ds}\right)^2-\left(\frac{dx}{ds}\right)^2=1 \,,
\end{eqnarray}
where $t=t(C_1,C_2,s)$, $x=x(C_1,C_2,s)$ are parametric equations of the trajectories in which any fixed values of the constants $C_1,C_2$ pick out a particular trajectory. 

\begin{figure}[ht]
	\begin{center}
		\includegraphics[width=3in]{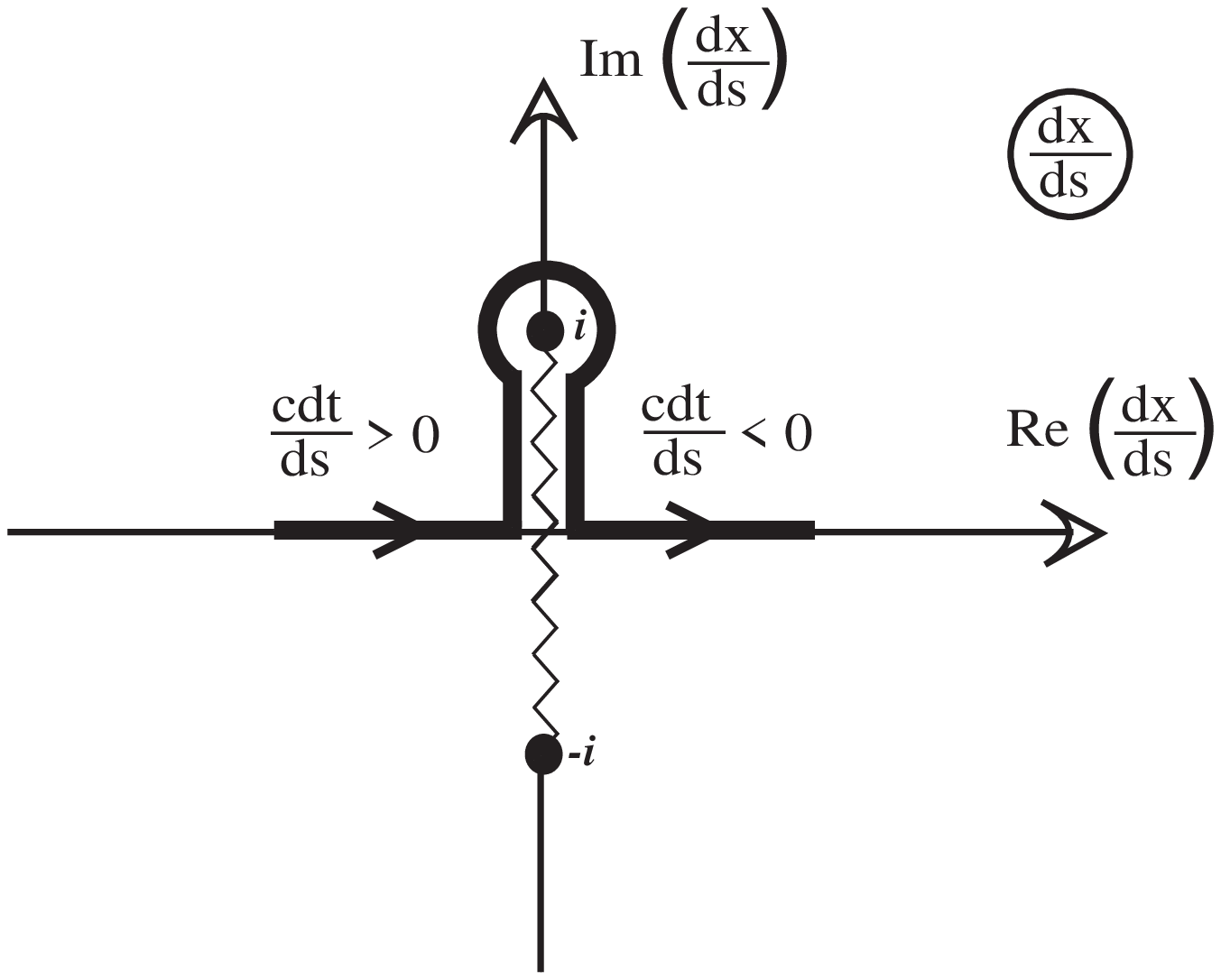}
	\end{center}
	\label{fig:2}
	\caption{}
\end{figure}

\begin{figure}[ht]
	\begin{center}
		\includegraphics[width=3in]{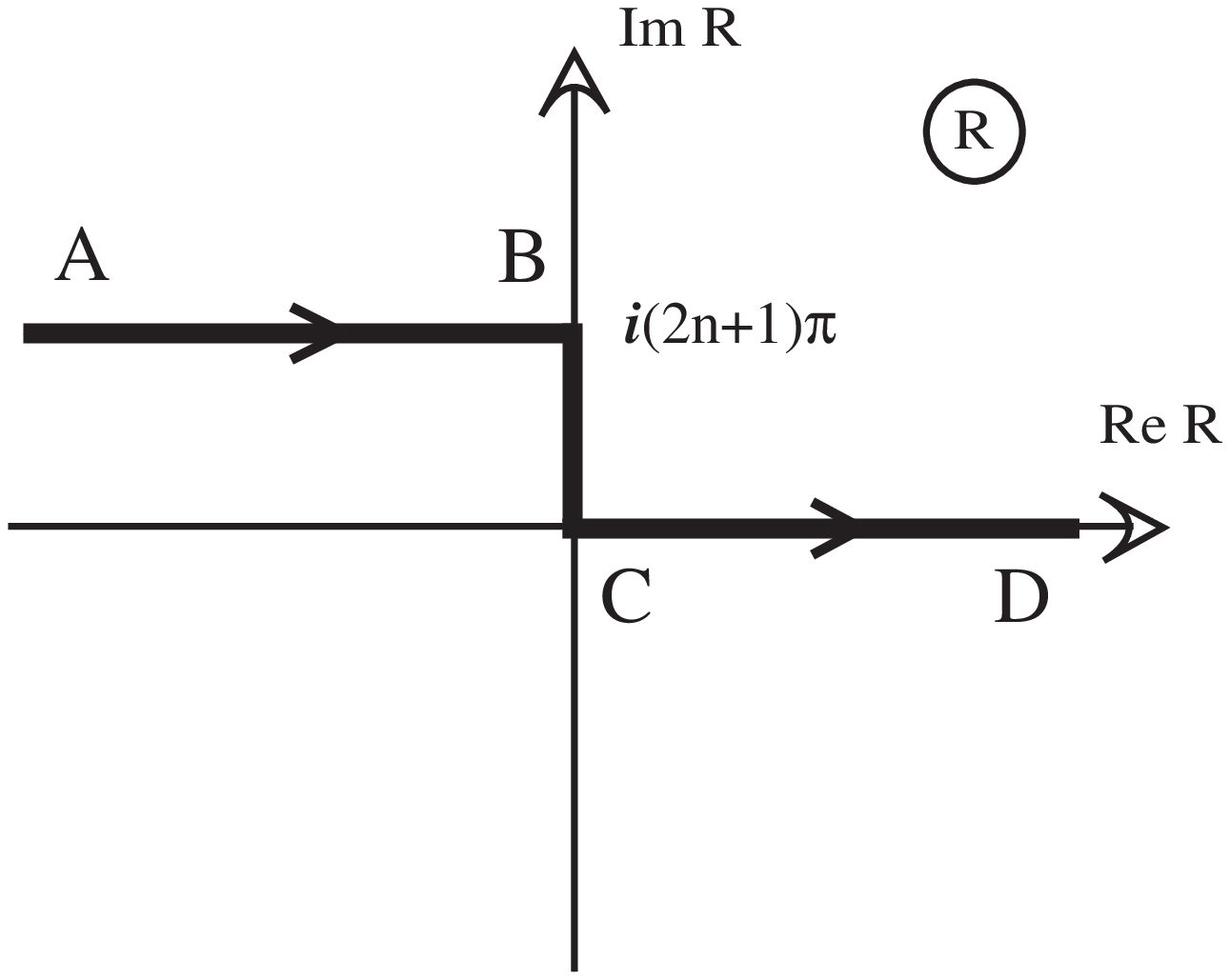}
	\end{center}
	\label{fig:3}
	\caption{}
\end{figure}

The quasiclassical path making the principal contribution to the pair creation probability corresponds to the continuous contour (shown in Fig.~2) in the complex plane of the variable $dx/ds$ containing the sub-barrier jump (vertical part of the contour) after which the quantity $cdt/ds$ changes sign.

Evidently such a path cannot be found for every external field. For example, in empty space $dx/ds$ is simply a constant so there does not exist any such trajectory. In a constant homogeneous electrical field $E$ directed towards increasing values of the coordinate $x$, the equations of motion of a particle with mass $m$ and charge $e$ are:
\begin{eqnarray}
	ct=\sigma \sinh\rho+C_1, \quad\quad x=\sigma\cosh\rho+C_2 \,,
\end{eqnarray}
where $\sigma=me^{-1}E^{-1}$ and $\rho=\sigma^{-1}s$. In this case the contour shown in Fig.~2 can be realized if the parameter $\rho$ in its own complex  plane will change from the left to the right along the horizontal line $\rho=i\pi+\textrm{Re}\,\rho$,($-\infty<\textrm{Re}\,\rho<0$) to the point $\rho=i\pi$, then down along the imaginary axis to the origin $\rho=0$, and then to the right along the real axis to infinity. In such a process one of the particles will be created at the point $(ct,x)=(C_1,C_2-\sigma)$ and the other at the point $(ct,x)=(C_1,C_2+\sigma)$.

The essential feature of this picture is that both the initial part of trajectory and its final part belong to the same real spacetime and correspond to real values of the coordinates $t,x$.

Together with the principal path, contributions of the highest order to the creation probability are coming from the trajectories making a finite number of turns around the cut $(i,-i)$ in the complex plane of the variable $dx/ds$ before they emerge on the other side of the barrier. An approach of this kind (the imaginary time method) in its application to the external electromagnetic field was developed in \cite{10n} where this method was used to reproduce exactly Schwinger's  results concerning particle production in a constant electric field.

The generalization of this method to the gravitational field is straightforward \cite{9}. For a particle moving in such a field we have the well known integral of the motion $g_{ik}(dx^i/ds)(dx^k/ds)=-1$ which can be written in the following form
\begin{eqnarray}
\label{intmot}
\left(l_i\frac{dx^i}{ds}\right)^2-\left(m_i\frac{dx^i}{ds}\right)^2=1
\end{eqnarray}
with the use of an orthonormal frame consisting of two vectors $l_i,m_i$ in terms of which the metric tensor can be represented as $g_{ik}=-l_i l_k+m_i m_k$, $(i,k=0,1)$. Here $l_i$ is a continuous future-directed timelike vector.

In order for particle creation to take place, there should exist a geodesic trajectory corresponding to the same type of complexified contour as in Fig.~2 but now in the complex plane of the variable $m_i(dx^i/ds)$. Globally such a trajectory can be described by the equations $x^0=x^0(C_1,C_2,s)$, $x^1=x^1(C_1,C_2,s)$ with some fixed values of the constants $C_1,C_2$ and its initial and final parts should represent an analytical  continuation of each other with respect to the proper time $s$. In addition both initial and final parts of such geodesics must belong to the real spacetime with real values of the coordinates $x^0,x^1$.
Finally,  the invariant $l_i(dx^i/ds)$ should change the sign after the sub-barrier transition.\footnote{The sign of the quantity $l_i(dx^i/ds)$ is invariant with respect to both coordinate transformations and Lorentz transformations of the frame, hence it represents a physical quantity with which one can distinguish between particle and antiparticle.}
Due to tunneling, the  parameter $s$ gains an extra imaginary part $\Delta s$. Because the particle's  action in the gravitational field is $-mc\int ds$, this imaginary part defines the order of magnitude of the main contribution to the probability of the process by means of the well known formula:
\begin{eqnarray}
\label{prob}
	w \sim \exp\left(-\frac{2m c}{\hbar}|\textrm{Im}\,\Delta s|\right) \,.
\end{eqnarray}
From the technical point of view, in place of $s$ it is more convenient to use the new parameter $R$ which can be defined by the equations:
\begin{eqnarray}
\label{Rpar}
	-l_i\frac{dx^i}{ds}=\cosh R, \quad\quad m_i\frac{dx^i}{ds}=\sinh R \,.
\end{eqnarray}
Without loss of generality we can think of the parameter $R$ as real on the final part of the trajectory and, if so, on its initial part this parameter can take only  the following complex values: $R=\textrm{Re}\, R+i(2n+1)\pi$, where $n$ is any integer including zero. This follows from the reality of the left hand sides of  Eqs.~(\ref{Rpar}) on the initial part of the trajectory and from the condition that the quantity $l_i(dx^i/ds)$ should have different signs on the initial and final parts of the trajectory. Then if a tunnel transition exists, it corresponds to the contour in the complex plane of $R$ which is shown in Fig.~3. The basic jump (which in the complex plane of the variable $m_i(dx^i/ds)$ is of the type depicted in Fig.~2) corresponds to the value $n=0$.  Nonzero positive values of $n$ are in keeping with higher corrections to the transition amplitude, when apart from the basic path the contour in complex plane $m_i(dx^i/ds)$ makes some finite number of turns around the cut $(i,-i)$.

The imaginary part of the difference of $R$ between the final and initial sectors of the trajectory is $\Delta R=i(2n+1)\pi$ and one can find $\textrm{Im}\,(\Delta s)$ if the relation between the parameters $R$ and $s$ is known. In this way one can get the main exponential factor in the probability for the process in accordance with the formula (\ref{prob}).

Now let us  apply  this approach to the metric (\ref{Krucoords}) taking the following frame:
\begin{eqnarray}
\label{newmet}
	l_i=(-\sqrt{\omega},0); \quad\quad m_i=(0,\sqrt{\omega}) \,,
\end{eqnarray}
where
\begin{eqnarray}
\label{newprob}
	\omega=\frac{4r_g^3}{r}e^{-\frac{r}{r_g}}\,.
\end{eqnarray}

The real spacetime with metric (\ref{Krucoords}) represents one physical sheet on which $r>0$ everywhere. Therefore we can define single-valued positive functions $\sqrt{r}$ and $\sqrt{\omega}$ in this region. Such a choice conforms to the condition that the timelike vector $l_i$ in (\ref{newmet}) should be everywhere directed into the future. Now the integral (\ref{intmot}) expressed through (\ref{Rpar}) takes the form:
\begin{eqnarray}
	\sqrt{\omega}\frac{d\eta}{ds}=\cosh R, \quad\quad \sqrt{\omega}\frac{d\zeta}{ds}=\sinh R
\end{eqnarray}
and it is easy to show that the geodesic equations for the metric (\ref{Krucoords}) allow one more integral:
\begin{eqnarray}
\label{newint}
	\omega\eta\frac{d\zeta}{ds}-\omega\zeta\frac{d\eta}{ds}=C_1 \,,
\end{eqnarray}
where $C_1$ is an arbitrary constant. From the last three formulas it follows that
\begin{eqnarray}
\left(\frac{dr}{ds}\right)^2=\frac{C_1^2-4r_g^2}{4r_g^2}+\frac{r_g}{r}
\end{eqnarray}
and it can be seen that the constant $C_1$ is connected with the conserved Schwarzschild energy ${\mathcal E}$ of the particle (with mass $m$)  by the equation:
\begin{eqnarray}
	C_1^2=\frac{4r_g^2{\mathcal E}^2}{m^2c^4} \,.
\end{eqnarray}
Most interesting for us is the case $C_1^2>4r_g^2$ corresponding to those  particles which can escape to infinity, so let us first study these kinds of geodesics. For such trajectories the solution of Eqs.~(\ref{newprob})--(\ref{newint}) can be expressed with the aid of the auxiliary parameter $\lambda$:
\begin{eqnarray}
\label{newzeta}
	\zeta=-\frac{C_1}{\sqrt{\omega}}\cosh R-\frac{2 r_g\alpha}{\sqrt{\omega}}\coth\lambda\sinh R \,,\\
	\eta=-\frac{C_1}{\sqrt{\omega}}\sinh R-\frac{2 r_g\alpha}{\sqrt{\omega}}\coth\lambda\cosh R \,,
\label{neweta}
\end{eqnarray}
where
\begin{eqnarray}
	\alpha=\frac{\sqrt{C_1^2-4r_g^2}}{2r_g}=\frac{\sqrt{{\mathcal E}^2-m^2c^4}}{mc^2}, \quad \alpha>0 \,.
\end{eqnarray}
The quantities $R$, $s$ and $r$ also can be expressed as functions of the same parameter $\lambda$:
\begin{eqnarray}
\label{C2}
	C_2&=&R+\frac{C_1}{8r_g\alpha^3}\left[\sinh 2\lambda +2(2\alpha^2-1)\lambda\right] \,,\\
\label{spar}
	s&=&\frac{r_g}{2\alpha^3}\left(\sinh 2\lambda-2\lambda\right) \,,\\
	\sqrt{r}&=&\frac{\sqrt{r_g}}{\alpha}\sinh\lambda \,,
\label{rpar}
\end{eqnarray}
where
in (\ref{C2}), $C_2$ is an arbitrary constant (yet another integral of motion). If the variable
$R$ changes along the contour ABCD (Fig.~3), we can see from (\ref{C2}) how the parameter $\lambda$ changes. For this we need to take into account the fact that on the branches AB and CD the quantity $\sqrt{r}$ is real and positive. It is always admissible to take all quantities to be real along the final branch CD and, if so,  from (\ref{rpar}) it follows that along AB (including the point B) and along CD (including the point C) the parameter $\lambda$ can be expressed as
\begin{eqnarray}
\label{lambdaAB}
	\lambda_{AB}=(\textrm{Re}\,\lambda)_{AB}+2iN\pi, \quad\quad \lambda_{CD}=(\textrm{Re}\,\lambda)_{CD} \,,
\end{eqnarray}
where $N$ is an integer number which remains to be calculated. Since $C_2$ is constant and does not change along the contour we should equate  the real and  imaginary parts of the right hand side of Eq.~(\ref{C2}) at the points B and C. From the equality of the real parts, taking into account (\ref{lambdaAB}) and the fact that the parameter $R$ is purely imaginary at the points B and C, one and only one result can be derived:
\begin{eqnarray}
	(\textrm{Re}\,\lambda)_B=(\textrm{Re}\,\lambda)_C \,.
\end{eqnarray}
From this and from (\ref{lambdaAB}) it follows that $\coth\lambda$ is real everywhere and
\begin{eqnarray}
	(\coth\lambda)_B=(\coth\lambda)_C \,.
\end{eqnarray}
Finally from Eqs.~(\ref{newzeta}) and (\ref{neweta}) we obtain:
\begin{eqnarray}
	(\eta,\zeta)_B=(-\eta,-\zeta)_C \,.
\end{eqnarray}
This is the result which confirms the our claim made at the beginning of this section. 

We should not forget the other case $C_1^2<4r_g^2$. This class of geodesics is of no interest, because even if such particles were able to be created they could not escape to spatial infinity. The situation, however, is even simpler: in this case particle creation is completely impossible. For such trajectories $\alpha$ is purely imaginary and all the equations can be deduced from (\ref{newzeta})--(\ref{rpar}) by the replacement $\lambda=i(\psi-\pi/2)$ keeping the parameter $R$ the same. The new auxiliary parameter $\psi$ must be real along final part CD of the contour ABCD and on its initial branch AB it must have an extra imaginary part proportional to the imaginary term $i(2n+1)\pi$ in the parameter $R$.  However, this contradicts the reality of $\sqrt{r}$ along the branch AB. Thus in this case a tunnel trajectory with both external parts belonging to the real spacetime\footnote{As has already been emphasized such a condition is indispensable for the existence of the pair creation effect. This condition has been ignored, however,  for the trajectories considered in \cite{11} as candidates for the tunneling paths.    

\hskip 20pt
Another example of an attempt to find quasiclassical tunneling for the problem at hand can be found in Ref.~\cite{12}.  This proposal is also irrelevant, because for a black hole to be able to radiate the corresponding sub-barrier transition \emph{must} exist  in its  \emph{fixed} external field which is not true for the approach taken there. Those authors also propose including  the black hole's  horizon in the quasiclassical dynamics, but for this one needs to develop first the full quantum theory of the system including quantization of the matter of the collapsing body. Whether such a theory can be constructed and whether the proposed quasiclassical dynamics can be derived from it is unclear. It is worth mentioning that the authors of Ref.~\cite{12} agree on the non-existence of  tunneling in the fixed black hole field (see the last page of Ref.~\cite{12}).} does not exist.

We make a few final comments about pair creation by the \emph{eternal} black hole. The reason for mentioning this is only for the formal completeness of the analysis since these objects have little to do with the physical black holes created during the collapse and this matter has no relation to the scope of the present paper. It seems that the probability of pair creation in the maximally extended Kruskal spacetime at the conjugate points $(\eta,\zeta)$ and $(-\eta,-\zeta)$ should be considered to be zero because the set of trajectories containing tunneling have zero measure with respect to the set of all trajectories. Indeed, equating the imaginary parts of the right hand side of Eq.~(\ref{C2}) at points B and C gives the following equation:
\begin{eqnarray}
\label{maineq}
	2n+1+\alpha^{-3}(2\alpha^2-1)\sqrt{1+\alpha^2}N=0
\end{eqnarray}
from which the number $N$ can be obtained if $2\alpha^2-1\neq0$. In the case when $2\alpha^2-1=0$ Eq.~(\ref{maineq}) cannot be satisfied no matter what the number $n$ is. This means that the trajectory corresponding to the  value $\alpha^2=1/2$ cannot contain a tunneling. For all other values of $\alpha$ the imaginary part of the change of the particle's  action under the transition ABCD can be found from (\ref{spar}) and (\ref{lambdaAB}):
\begin{eqnarray}
	\textrm{Im}\,(\Delta s)=\alpha^{-3}r_g\textrm{Im}\,(\lambda_{AB}-\lambda_{CD})=2\alpha^{-3}r_g\pi N \,.
\end{eqnarray}
It seems that by substituting the number $N$ which follows from (\ref{maineq}) into this expression, one can evaluate the probability of tunneling according to the formula (\ref{prob}). For the main contribution, that is for $n=0$, one gets:
\begin{eqnarray}
	w\sim\exp\left(\frac{-4\pi mc r_g}{\hbar\sqrt{1+\alpha^2}|1-2\alpha^2|}\right) \,.
\end{eqnarray}
The trouble is that the constant $\alpha$ comes from Eq.~(\ref{maineq}) in which  both numbers $N$ and $n$
are integers. Then those trajectories with the ability to open a path through the barrier correspond to a countable  set (that is a set of the same size as the rational numbers) of the values of this constant. Then the measure of the set of such trajectories in any interval of $\alpha$ should be evaluated as zero. Consequently the total probability of the tunneling in any interval of $\alpha$ also should be evaluated as zero. [Nothing like this can be seen for particle production in a constant electric field; there each trajectory, independently of the values of  constants of motion, has the potential ability to tunnel through the barrier, that is the  measure of the set of the such trajectories  is just unity.] Besides, the sense of the phenomenon discussed also remains unclear due to the following reason. If a quasiclassical approach demonstrates the absence of tunneling trajectories, the question can be considered as solved. In the case of its presence, the answer is still unclear because in addition one needs to be sure that an exact theory exists from which that quasiclassical result can be deduced in some appropriate limit. The last requirement, however, is problematic for the external fields without in-  and out- regions where a sensible notion of particles can be defined. 

\section{The second quantization approach}

In the conventional approach one assumes that the following modes of positive Schwarzschild frequency in the external region $r>r_g$ and  outside of the shell (sector CDH, Fig.~1) are related to the particles escaping to infinity:
\begin{eqnarray}
\label{phimodes}
	\varphi_\mu=A_\mu e^{-i\mu(ct-r^*)}\,, \quad\quad \mu>0 \,,
\end{eqnarray}
where $A_\mu$ are the normalization constants. These modes are solutions of the massless Klein-Gordon equation for the metric (\ref{intout}). Using the equations of motion of the shell (\ref{eqm}) and relation (\ref{corr}) between the times $t$ and $\tau$, it is easy to find these modes inside the shell. To do this we need to match solutions on the shell and to take into account the zero boundary condition for the field at the points $r=0$ (the line AEF, Fig.~1). This condition can be used to mimic the total wave  reflection from the center of the collapsing body. In this way it can be seen that the modes (\ref{phimodes}) arise due to the reflection from the center of those waves coming from infinity in the region $c\tau+r<0$ (that is in triangle EFG, Fig.~1) and having in this region the following form:
\begin{eqnarray}
\label{phimodesref}
	\varphi_\mu=-A_\mu\exp\left\{-2i\mu r_g[v-\ln(-v)]\right\} \,, \quad\quad v<0 \,,
\end{eqnarray}
where
\begin{eqnarray}
	v=\frac{c\tau+r}{2r_g}\,.
\end{eqnarray}
We see that these modes are singular on the last ray $v=0$ (line EG, Fig.~1) and in a traditional description just this singularity is responsible for the outgoing late time flux. Since we are interested only in this asymptotically late radiation, we can simplify the calculations further, neglecting the linear term (in $v$) with respect to the logarithm in (\ref{phimodesref}) not only near the point $v=0$ but  everywhere on the $v$-axis (this is a standard simplification in the conventional approach, see, for example, Ref.~\cite{13}). Then for the incoming modes we let
\begin{eqnarray}
\label{modesinc}
	\varphi_\mu=-A_\mu\exp\left[2i\mu r_g\ln(-v)\right] \,, \quad\quad v<0 \,, \quad \mu>0 \,.
\end{eqnarray}
Besides, in the conventional treatment the shape of the modes for $v>0$ (to the right from the last ray EG, Fig.~1) is of no importance, since these waves are not able to escape to infinity, and therefore the late time flux cannot depend on its concrete form \cite{13}, \cite{14}.  In particular, it is considered that the modes for $v>0$ can be put simply to zero, that is the modes responsible for the effect are of the form:
\begin{equation}
\label{modes}
L_\mu=\left\{
	\begin{array}{l}
		-A_\mu\exp\left[2i\mu r_g\ln(-v)\right] \,, \quad\quad v<0 \,, \quad \mu>0 \,, \\
		0 \,, \quad\quad v>0 \,.
	\end{array}
\right.
\end{equation}
Nevertheless for the quantization of the field at the initial stages of the collapse we need a complete basis, that is we need to choose the incident modes in some appropriate nontrivial form also in the region $v>0$. In the traditional approach a convenient choice is \cite{13}, \cite{14}:
\begin{equation}
\label{tradmodes}
R_\mu=\left\{
	\begin{array}{l}
		-A_\mu^*\exp\left[-2i\mu r_g\ln(v)\right] \,, \quad\quad v>0 \,, \quad \mu>0 \,, \\
		0 \,, \quad\quad v<0 \,.
	\end{array}
\right.
\end{equation}
Here and below we use an asterisk to denote complex conjugation (only in the tortoise coordinate $r^\ast$ above does the asterisk not have this interpretation, used there since is the conventional symbol). Later it will become clear that this choice of modes is  adequate to the problem we are interested in.

At the beginning of the collapse the shell's radius is infinite and the whole spacetime is flat with metric (\ref{intin}) both inside and outside the shell. Consequently the field in this in-region is free and its incident part can be quantized in the standard way using plane wave modes $\psi_\omega$:
\begin{eqnarray}
	\psi_\omega=(4\pi\omega)^{-1/2}\exp(-2i\omega r_g v) \,, \quad\quad \omega>0 \,, \quad -\infty<v<\infty \,.
\end{eqnarray}
These modes should be taken in the standard Minkowski vacuum state in accordance with the initial condition that no particles are coming in from infinity. 

Now to solve the problem of late time particle production one needs to compare this standard plane wave quantization  of the incident field with its quantization on modes (\ref{modes}) and (\ref{tradmodes}) and find the relation between ladder operators in both quantization schemes. Naturally, this can be done only if the set of the modes (\ref{modes}) and (\ref{tradmodes}) is complete and this is the basic assumption of the conventional approach. In other words the assumption is that the complete set of the singular falling modes of the type (\ref{modesinc}) at the initial stages of collapse (that is in Minkowski spacetime) can be represented by two sets: the ``left" modes $L_\mu$ which are nonzero only to the left of the last ray $(v<0)$  and ``right" modes $R_\mu$ which are nonzero only to the right of the last ray $(v>0)$. Both sets are characterized by positive values of the quantum number $\mu$. At the same time all left modes are orthogonal to each right one and vice versa simply because these sets have supports in different nonintersecting regions of the spacetime. In addition the coefficients $A_\mu$ in (\ref{modes}), (\ref{tradmodes}) can be defined in such a way that the modes of each set will be orthonormal among themselves. Orthonormality here is intended to be with respect to the scalar product in the initial Minkowski spacetime with metric (\ref{intin}):
\begin{eqnarray}
\label{modesdef}	(\varphi,\psi)=i\int_{-\infty}^{\infty}\left(\varphi^*\frac{\partial\psi}{\partial  v}-\psi\frac{\partial\varphi^*}{\partial v}\right)dv \,.
\end{eqnarray}
In accordance with the assumed completeness of the left and right modes, the field can be decomposed in terms of these modes and this expansion can be equated to the plane wave mode expansion:
\begin{eqnarray}
\label{expan}
	\int_0^\infty\left(l_\mu L_\mu+r_\mu R_\mu+l_\mu^\dagger L_\mu^*+r_\mu^\dagger R_\mu^*\right)d\mu=\int_0^\infty\left(a_\omega\psi_\omega+a_\omega^\dagger \psi_\omega^* \right)d\omega \,.
\end{eqnarray}
In virtue of the orthonormality of the left and right modes we obtain
\begin{eqnarray}
\label{ort}
l_\mu=\int_0^\infty\left[a_\omega(L_\mu,\psi_\omega)+a_\omega^\dagger(L_\mu,\psi_\omega^*)\right]d\omega \,.
\end{eqnarray}
It is easy to check that $(L_\mu,\psi_\omega^*)\neq0$ which implies
\begin{eqnarray}
\label{cond1}
	\bra{O_M}l_\mu^\dagger l_\mu\ket{O_M}\neq0 \,,
\end{eqnarray}
where $\ket{O_M}$ is the vacuum with respect to the operators $a_\omega$, that is the state which coincides with the Minkowski vacuum at the initial stages of the collapse. Not only is the quantity (\ref{cond1}) nonzero but it is divergent and this result has  the traditional interpretation as corresponding to the permanent production of the left particles, namely those creating the stationary flux detected by the distant Schwarzschild observer in the far future.

We see also that in the approach being discussed, the right modes indeed are of no significance since they contribute nothing to this flux.

However, more thorough analysis shows that the scheme we have just described is insolvable because the set of the left and right modes is incomplete for free incident waves in Minkowski spacetime and such waves cannot be represented as a linear superposition of these modes. This means that the equality (\ref{expan}) in fact does not exist, and hence the relation (\ref{ort}) between ladder operators also does not exist. Then for the real physical case, the expression (\ref{cond1}) for the particle number has no sense independent of whether it is finite or divergent.

Let us look more carefully into how one can construct a complete set of modes of the type (\ref{modesinc}) for the incident waves at the early phases of collapse. To do this it is necessary to continue analytically the solutions (\ref{modesinc}) into the region $v>0$ through the lower half-plane of the complex variable $v$ and permit the quantum number $\mu$ to take all real values. The latter is due to the fact that with respect to the time $\tau$ at the beginning of the collapse, the number $\mu$ has nothing to do with frequency and it is not required to only be positive. It is more convenient to designate this extension of the number $\mu$ by another letter, for instance $\kappa$, and preserve for $\mu$ only positive values keeping in mind that for $\kappa>0$ the numbers $\kappa$ and $\mu$ coincide identically. These continuations gives the following globally defined modes:
\begin{equation}
\label{globmodes}
{\mathcal L}_\kappa=\left\{
	\begin{array}{l}
		B_\kappa e^{r_g\pi\kappa}\exp\left[2i r_g\kappa\ln(-v)\right] \,, \quad\quad v<0 \,,\\
		B_\kappa e^{-r_g\pi\kappa}\exp\left[2i r_g\kappa\ln(v)\right] \,, \quad\quad v>0 \,,
	\end{array}
\right.
\end{equation}
where $-\infty<\kappa<\infty$ and $B_\kappa$ are a new constants which can be found from the orthonormality conditions for the modes ${\mathcal L}_\kappa$ with respect to the scalar product (\ref{modesdef}). These conditions are
\begin{eqnarray}
\label{cond2}
	({\mathcal L}_\kappa,{\mathcal L}_{\kappa'})=\delta(\kappa-\kappa') \,, \quad\quad 
({\mathcal L}_\kappa,{\mathcal L}_{\kappa'}^*)=0
\end{eqnarray}
and calculations show that
\begin{eqnarray}
\label{Bpar}
	B_\kappa=\frac{\sqrt{r_g}}{2\pi}\Gamma(-2ir_g\kappa) \,,
\end{eqnarray}
where $\Gamma$ is the standard gamma function. From the same calculations it follows that modes ${\mathcal L}_\kappa$ have the integral representation
\begin{eqnarray}
\label{intrep}
	{\mathcal L}_\kappa=\frac{\sqrt{r_g}}{2\pi}\int_{-\infty}^\infty d\gamma\exp(-iv e^\gamma-2ir_g\kappa\gamma) \,,
\end{eqnarray}
where the quantities $v$ and $\kappa$ should be understood as $v-i0$ and
$\kappa+i0$ respectively. It can be  checked  (see \cite{15}, for example)  that the integral (\ref{intrep}) indeed gives all three results (\ref{globmodes})--(\ref{Bpar}).

From the representation (\ref{intrep}) it is easy to see that ${\mathcal L}_\kappa$ for any value of $\kappa$ are the positive-frequency modes for the time $\tau$ and vice-versa, any positive-frequency plane wave has an expansion in terms of the modes ${\mathcal L}_\kappa$, since the formula (\ref{intrep}) can be inverted easily:
\begin{eqnarray}
	\exp(-iv e^\gamma)=2\sqrt{r_g}\int_{-\infty}^\infty {\mathcal L}_\kappa e^{2i r_g\kappa\gamma}d\kappa \,.
\end{eqnarray}
Because the plane-wave set is complete, from this invertibility completeness also follows for the set of modes ${\mathcal L}_\kappa$. 

All these calculations are simple and are based on well-known results. Formulas (\ref{globmodes}), (\ref{Bpar}) and (\ref{intrep}) can be found in many handbooks (\cite{15} is one of them) 
and the modes ${\mathcal L}_\kappa$ have been exhibited already in \cite{16} and later in \cite{17}. These modes can be obtained also from the so-called boost modes \cite{1},\cite{2},\cite{4} (Minkowski Bessel modes in the Gerlach terminology \cite{17}) for the massive scalar field in Minkowski spacetime by an appropriate limit to the massless case.

The foregoing properties of the modes ${\mathcal L}_\kappa$ mean that the quantization 
\begin{eqnarray}
\label{quant}
	\Phi=\int_{-\infty}^\infty(b_\kappa{\mathcal L}_\kappa+b_\kappa^\dagger{\mathcal L}_\kappa^*)d\kappa \\
	\left[b_\kappa,b_{\kappa'}^\dagger\right]=\delta(\kappa-\kappa'), \quad\quad \left[b_\kappa,b_{\kappa'}\right]=\left[b_{\kappa}^\dagger,b_{\kappa'}^\dagger\right]=0
\end{eqnarray}
is unitarily equivalent to the usual quantization in the plane wave basis and the $b$-vacuum remains the same as the Minkowski vacuum:
\begin{eqnarray}
	b_\kappa\ket{0_M}=0 \,.
\end{eqnarray}
It is important that with respect to the variable $\kappa$, the modes ${\mathcal L}_\kappa$ must be treated as distributions considering $v$ as parameter \cite{18,4}. Therefore these modes are well defined for all values of the coordinate $v$ including $v=0$.  It can be seen from (\ref{intrep}) that on the last ray, that is for $v=0$, the modes ${\mathcal L}_\kappa$ are concentrated on the zero value of spectral parameter $\kappa$:
\begin{eqnarray}
\label{zeromodes}
	{\mathcal L}_\kappa=\frac{1}{2\sqrt{r_g}}\delta(\kappa) \,, \quad\quad v=0 \,,
\end{eqnarray}
where $\delta(\kappa)$ is the standard delta-function. This means that the integration interval over $\kappa$ in (\ref{quant}) in principle cannot be split into two parts: over negative and positive values of $\kappa$. Having made such  a splitting we lose an essential part of the degrees of freedom of the quantum field, namely those concentrated at the spectral point $\kappa=0$ and represented by the delta-function (\ref{zeromodes}). Such an operation is equivalent to the replacement of the initial system (that is a free field in  Minkowski space) by  another one which  has quite different physical behaviour in the vicinity of the last ray. But it is exactly this operation that the conventional approach is based on. The quantization in the left and right mode basis (\ref{modes})--(\ref{tradmodes}) means that the quantization is over an incomplete mode set which can be constructed only after excluding from the system the singular degree of freedom (\ref{zeromodes}). Indeed, after such an exclusion the expansion (\ref{quant}) can be written as
\begin{equation}
\begin{array}{rcl}
\label{newexpan}
	\tilde\Phi &=& \displaystyle{\int_{-\infty}^0}(\tilde b_\kappa {\mathcal L}_\kappa+\tilde b_\kappa^\dagger {\mathcal L}_\kappa^*)d\kappa+\int_{0}^\infty(\tilde b_\kappa {\mathcal L}_\kappa+\tilde b_\kappa^\dagger {\mathcal L}_\kappa^*)d\kappa \,, \\
	&=& \displaystyle{\int_0^\infty}(\tilde b_\mu {\mathcal L}_\mu+\tilde b_{-\mu} {\mathcal L}_{-\mu}+\tilde b_\mu^\dagger {\mathcal L}_\mu^*+\tilde b_{-\mu}^\dagger {\mathcal L}_{-\mu}^*)d\mu \,.
\end{array}
\end{equation}
Here the tilde indicates that after we remove the quantum fluctuations corresponding to the singular spectral point $\kappa=0$, the field and creation-destruction operators are different from those in (\ref{quant}). Now it is easy to see that the expansion (\ref{newexpan})  can be represented as an expansion over left and right modes. The latter have the following expression in terms of the modes ${\mathcal L}_\kappa$:
\begin{equation}
\begin{array}{rcl}
\label{lrmodes}
	L_\mu&=&\displaystyle{\frac{1}{\sqrt{2\sinh(2\pi r_g \mu)}}}\left(e^{\pi r_g \mu}{\mathcal L}_\mu-e^{-\pi r_g \mu}{\mathcal L}_{-\mu}^*\right) \,, \quad\quad \mu>0 \,, \nonumber\\
	R_\mu&=&\displaystyle{\frac{1}{\sqrt{2\sinh(2\pi r_g \mu)}}}\left(e^{\pi r_g \mu}{\mathcal L}_{-\mu}-e^{-\pi r_g \mu}{\mathcal L}_\mu^*\right), \quad\quad \mu>0 \,.
\end{array}
\end{equation}
The substitution of expressions (\ref{globmodes}) and (\ref{Bpar}) for the modes ${\mathcal L}_\kappa$ into (\ref{lrmodes}) gives exactly (\ref{modes}) and (\ref{tradmodes}) in which
\begin{eqnarray}
	A_\mu=-\frac{\sqrt{r_g\sinh(2\pi r_g\mu)}}{\pi\sqrt{2}}\Gamma(-2i r_g\mu) \,.
\end{eqnarray}
Using  (\ref{lrmodes}) and (\ref{cond2}) it is easy to show that for $\mu>0$ and $\mu'>0$, the orthonormality relations for the left and right modes are exactly those needed for the traditional picture:
\begin{eqnarray}
	(L_\mu,L_{\mu'})&=&\delta(\mu-\mu') \,, \quad\quad (L_\mu,L_{\mu'}^*)=0 \,, \nonumber\\
	(R_\mu,R_{\mu'})&=&\delta(\mu-\mu') \,, \quad\quad (R_\mu,R_{\mu'}^*)=0 \,, \\
	(L_\mu,R_{\mu'})&=&(L_\mu,R_{\mu'}^*)=0 \,. \nonumber
\end{eqnarray}
If we invert the formulas (\ref{lrmodes}) and substitute the results into the expansion (\ref{newexpan}) we get:
\begin{eqnarray}
	\tilde\Phi=\int_0^\infty(l_\mu L_\mu+r_\mu R_\mu+l_\mu^\dagger L_\mu^*+r_\mu^\dagger R_\mu^*)d\mu \,,
\end{eqnarray}
where
\begin{equation}
\begin{array}{l}
	l_\mu=\tilde b_\mu\cosh \theta_\mu+\tilde b_{-\mu}^\dagger\sinh\theta_\mu \,,
	r_\mu=\tilde b_{-\mu}\cosh \theta_\mu+\tilde b_\mu^\dagger\sinh\theta_\mu \,,
\end{array}
\end{equation}
and parameters $\theta_\mu$ are defined by the equation:
\begin{eqnarray}
	\tanh\theta_\mu=e^{-2\pi r_g\mu} \,, \quad\quad \mu>0 \,.
\end{eqnarray}
The $\tilde b$-vacuum state $\ket{\tilde 0}$ satisfies the conditions
\begin{eqnarray}
	\tilde b_\mu\ket{\tilde 0}=0 \,, \quad\quad \tilde b_{-\mu}\ket{\tilde 0}=0 
\end{eqnarray}
and its physical interpretation is known \cite{10}. This state is a condensate, containing an infinite number of pairs of real particles. Each pair consists from one left and one right particle (corresponding to the $l$ and $r$ operators) correlated to each other. To each value of $\mu$ corresponds an infinite number of the left particles with finite density: 
\begin{eqnarray}
\label{stat}
	\displaystyle{(n_\mu)_l=\sinh^2\theta_\mu=\frac{1}{e^{4\pi r_g \mu}-1}} \,.
\end{eqnarray}
The quantity $(n_\mu)_l$ is the density of the expectation value of the operator $l_\mu^\dagger l_\mu$ in the vacuum $\ket{\tilde 0}$.

At the initial stages of the collapse, expression (\ref{stat}) has no meaning as a physical Bose distribution because $\mu$ is not the particle's energy. However, after the left particles pass the collapsing body and are registered at late time by an external observer, the latter indeed can say that the density of the detected particles corresponds to an ideal Bose gas at the temperature
\begin{eqnarray}
	T=\frac{\hbar c^3}{8\pi k G M}
\end{eqnarray}
because $r_g=\frac{2 G M}{c^2}$ and from the observer's  point of view,  the particle's  energy is $\mu \hbar c$ ($G$ is the Newtonian gravitational constant and $k$ is Boltzmann constant).

However, as already emphasized above, this picture has nothing in common with the real problem when collapse starts in the standard Minkowski vacuum because the initial state $\ket{\tilde 0}$ in no sense can be considered to be the Minkowski vacuum. The left particles falling into the system from infinity are already present in the initial state $\ket{\tilde 0}$ and they  are detected at late time after passing through the collapsing matter. Also the incident right particles are reflected by the collapsing body and appear inside the horizon (in the region ABCD, Fig.~1). Then near the horizon DC we really have pairs of correlated particles, one outside (moving to spatial infinity)  and one inside (falling into the singularity) but they originated from the singular (on the last ray) incident flux  which already existed in the state $\ket{\tilde 0}$ in the infinite past. There does not exist any late time process of particle production. 
   
It is reasonable to ask whether one can  find some physically sensible model in which the initial incident waves could be in the state $\ket{\tilde 0}$. The answer is rather negative because all such models would be unavoidably too exotic. One of the possibilities is to cut the last ray out of the spacetime, that is to prepare a new  manifold without those points where the modes ${\mathcal L}_\kappa$ are concentrated at one spectral value $\kappa=0$. In this case the spectral point $\kappa=0$ will be a regular point for the functions ${\mathcal L}_\kappa$ everywhere in the spacetime and the splitting (\ref{newexpan}) will become permissible. Naturally, such a model has nothing to do with the real one because by cutting away the last ray we essentially change the topology of the spacetime. Now we are dealing with two universes without common points and the construction is nothing else but a realization of the known formal procedure of the doubling of the degrees of freedom of the system in order to describe the mixed state in the left  universe  by the pure state ($\tilde b - \hbox{vacuum}$) in the doubled world. It is evident that in this case the Minkowski vacuum is out of the question simply because we have no Minkowski spacetime.

Another interpretation is that we are in Minkowski spacetime but with the zero boundary condition for the incident waves at the points of the last ray. In this case the set of the left and right modes will be complete (in the sense of distributions, these modes vanish on the last ray) and the traditional quantization procedure can be applied. In such a model the space of the states also cannot contain the Minkowski vacuum since the latter is translationally invariant and this property is incompatible with such a physically special status of the last ray. 

A third way of thinking would be not to exclude the mode concentrated at $\kappa=0$
from the system but only from that part of the field which is going to be quantized. This means that we consider the state with $\kappa=0$ to be filled by the large number of particles which make up a classical condensate and on its background the quantum fluctuations of the ``zero mode" are ignorable. Such model is an analogue of the one proposed in \cite{6} in relation to the quantization of the massive scalar field over the left and right Unruh modes in Minkowski spacetime. Needless to say in this case the initial state of the incident field is also not  the Minkowski vacuum. 

\section*{Acknowledgments}

I would like to thank H. Kleinert for his interest to this work and stimulating discussions and R. Jantzen for the improvement of English.

\end{document}